# Exchange bias in laterally oxidized Au/Co/Au nanopillars

Ll. Balcells[a#], B. Martinez[a], O. Iglesias[b] J.M. García-Martín[c], A. Cebollada[c], A. García-Martín[c], G. Armelles[c], B. Sepúlveda[d] and Y. Alaverdyan[e]

a) Institut de Ciència de Materials de Barcelona (ICMAB-CSIC), 08193 Bellaterra, Spain.
b) Departament de Física Fonamental and Institut de Nanociència i Nanotecnologia, Universitat de Barcelona, Martí i Franquès 1, 08028 Barcelona, Spain.
c) Instituto de Microelectrónica de Madrid, IMM (CNM-CSIC), Tres Cantos, Spain.
d) Research Center of Nanoscience and Nanotechnology (CIN2-ICN/CSIC), 08193 Bellaterra, Spain.
e) Cavendish laboratory, University of Cambridge, CB3 0HE Cambridge, United Kingdom.

**Abstract**

Au/Co/Au nanopillars fabricated by colloidal lithography of continuous trilayers exhibit and enhanced coercive field and the appearance of an exchange bias field with respect to the continuous layers. This is attributed to the lateral oxidation of the Co interlayer that appears upon disc fabrication. The dependence of the exchange bias field on the Co nanodots size and on the oxidation degree is analyzed and its microscopic origin clarified by means of Monte Carlo simulations based on a model of a cylindrical dot with lateral core/shell structure.

# Corresponding e-mail: balcells@icmab.es

Exchange bias (EB) arises from direct exchange interactions at the interface between ferromagnetic (FM) and antiferromagnetic (AF) systems [1,2]. It generates a unidirectional anisotropy which precludes the reversal of FM moments leading to an enhancement of the coercivity and a shift of the hysteresis loops along the field axis after field cooling. Since EB may be useful for stabilizing magnetic moments of small FM particles [3], there is an increasing interest in clarifying its mechanisms and optimizing its effects. The relevance of the AF/FM interfacial microstructural quality and the existence of a thickness threshold of the AF layer [4,5] for generating EB have been studied recently. In the case of Co nanoparticles the formation of a few nm thick antiferromagnetic (AF) CoO layer on the surface due to oxygen exposure promotes very often the appearance of a EB field that may drastically modify the magnetic behavior of the system [3,4, 6,7]. For this reason core/shell structured nanoparticles have been a subject of intense research.

In this paper, we analyze the dependence of the EB field on the particle size and on the oxidation degree in a nanostructured network of Co nanodots fabricated by using colloidal lithography (CL) [8].

Samples used in this work have been fabricated from metal trilayer films composed of 6 nm of Au, 10 nm of Co and 16 nm of Au sputtered onto glass substrates by using CL (see Ref. [9] for details). To obtain Co nanodots of different sizes latex spheres with different diameters (namely 60, 76 and 110 nm) (corresponding to samples N60, N76 and N110) have been used. This system exhibits very interesting surface plasmon resonance induced magneto-optical properties reported elsewhere [9]. In Fig. 1 we show an atomic force microscopy (AFM) image corresponding to sample N60. Co nanodots have the shape of truncated cones due to the leftovers of the polystyrene spheres. Thus, nanodots are made of a flat sandwich of 10 nm thick Co disk between two gold disks. From the AFM images the occupancy has been estimated to be around 25% of the total film. Magnetic properties were studied by using SQUID magnetometry and polar Kerr effect. Kerr effect measurements at room temperature indicate that: the dots are FM, the magnetization is in the film plane in both continuous and nanostructured samples, and only a small difference in the saturation magnetic field, $H_S$, has been observed. Magnetic measurements show that Co nanodots are FM at room temperature. As expected, the coercive field, $H_C$, is higher in nanostructured samples (see Fig. 2a), reflecting the increase of coercivity due to the reduction of particle size generated by the nanostructuration process. Nevertheless, no significant changes of the saturation magnetization ($M_S$) have been observed. Interestingly, at low temperature, a strong enhancement of $H_C$ is detected in nanostructured samples in parallel with the appearance of EB not observed in continuous films (see Fig.2b). To determine the origin of the EB field, we have analyzed the samples by using X-ray photoemission spectroscopy (XPS). We show the XPS spectrum for one of the samples in Fig. 3. After Ar ion milling 4 out of the 6 nm of the topmost Au layer, only peaks corresponding to metallic Co are found in non-nanostructured areas. In contrast, in nanostructured zones with Au/Co/Au nanopillars typical peaks corresponding to $Co^{2+}$ of CoO are detected. These peaks, corresponding to $Co^{2+}$, disappear with further $Ar^+$ ion milling, recovering the peaks corresponding to metallic Co. This indicates that the Au capping is effective in protecting the Co layer against oxidation in continuous and the patterned films.

Nevertheless, the nanostructuration process leads to the exposure to oxygen of lateral areas of Co disks in the Au/Co/Au nanopillars. Thus, the observed EB in nanostructured samples is likely to be generated by a very thin CoO layer (from $Ar^+$ ion milling process we estimate this layer is ~ 1 nm thick, which is less than 10% of the total Co volume of the nanodot) produced by oxidation of the lateral area of Co nanodots exposed to oxygen (see inset of Fig. 3). Using the simple Meiklejohn-Bean model [10] the critical thickness, $t^C_{AF}$, for generating a EB field is given by $t^C_{AF}=(J_{AF/F}S_{AF}S_F)/K_{AF}$, where $J_{AF/F}$ is the exchange interaction across the interface between the AF and FM layers, $S_{AF}$ and $S_F$ are respectively the spin of the AF and FM atoms at the interface and $K_{AF}$ is the anisotropy constant of the AF layer. Using the values given in [11] for the different constants, a value of $t^C_{AF} \approx 1nm$ is obtained for the CoO/Co system in agreement with experimental estimations previously reported [12]. To check whether further oxidation of the Co nanodots modifies EB, we have submitted a nanostructured sample to an oxidation process in ozone (N60oz). A clear enhancement of the EB field is observed (see Fig. 2b). As shown in Fig. 2b, $H_{EB}$ for N60 sample is almost four times larger after ozone attack ($H_{EB} \approx 200$ Oe after and $H_{EB} \approx 55$ Oe before) without no significant change of $H_C$ and only a small decrease of $M_S$. Hysteresis loops for the three samples (N60, N76 and N110) with different particle sizes are shown in Fig. 2c. The $H_c$ is slightly reduced when increasing the dot size as expected. It is found that $H_{EB}$ does not exhibit strong variations with dot size, thus indicating that EB can be induced by a thin oxidized lateral shell even for dots as big as 110 nm.

To better understand the microscopic mechanism responsible for the observation of EB in our system, we have performed atomistic Monte Carlo simulations based on a model in which dots are represented by cylinders of height H and diameter D cut out from a simple cubic lattice of Heisenberg spins that represent the Co ions. The spins interact through a Hamiltonian similar to that used for the simulation of core/shell magnetic nanoparticles in Ref. [13]

$$H = -\sum_{\langle i,j \rangle} J_{ij} \left( \vec{S}_i \cdot \vec{S}_j \right) - \sum_i k_i \left( \vec{S}_i \cdot \hat{n} \right)^2 - h \sum_i \hat{n} \cdot \vec{S}_i$$

The spins inside a shell of thickness $t_{Sh}$ around the lateral surface of the cylinder correspond to CoO and, therefore, are considered to interact AF ($J_{ij} = J_{Sh} < 0$), those inside the core of diameter $D - t_{Sh}$ have FM ($J_{ij} = J_C > 0$) interactions, while for those at the core/shell interface are coupled AF through $J_{ij} = J_{Int} < 0$). The anisotropy is uniaxial pointing along the in-plane field direction $\hat{n}$, and the anisotropy constants for core and shell spins have values $k_C = 0.022$ K, $k_{Sh} = 35$ K typical of bulk Co and CoO, h is the magnetic field per unit magnetic moment in temperature units [14]. For the dimensions of the studied structures, the number of spins in a dot considered in the simulation becomes prohibitively large for the present computer capabilities. Therefore, we adopted a recently proposed scaling approach [15] for the simulation of the magnetic configuration of nanostructures that has been successfully applied also to EB dots [16]. Accordingly, if the exchange interactions [14] are scaled by a factor $x = 0.104$, the dimensions of the 60 nm dot can be scaled by a factor $x^\eta = 0.285$ with $\eta \simeq 0.55$, resulting in a cylinder of reduced dimensions $D' = 48a$ and $H' = 8a$ (where $a = 3.55$ Å is the lattice constant) with $N = 13224$ spins, 17% of which are in the shell for $t_{Sh} = 3a$.

The simulated hysteresis loops were obtained after cooling at a temperature $T = 0.1$ K from a temperature higher than $T_C$ in a field $h_{FC} = 2$ K and following a protocol similar to that described in Ref. [13]. In Fig. 4a, the results of the simulation for a dot with AF shell of $t_{Sh} = 3a$ (red symbols) are shown, together with a dot with no shell (dashed line).

The hysteresis loop of the oxidized dot is shifted by $h_{EB} \simeq -0.1$ K towards the FC direction as in the experiment, demonstrating that a thin oxidized shell surrounding the lateral surface of the dot is able to produce an appreciable EB. The shape of the simulated loop is more squared than the experimental one due the fact that we are simulating an individual particle. The loop also presents a slight asymmetry between the decreasing and increasing field branches that can be attributed to different magnetization reversal processes, as can be inferred in the snapshots of the magnetic configurations included in the insets of Fig 4a, taken at the respective coercive field points. The coercive field of the oxidized dot ($h_C \simeq 0.95$ K) is increased with respect to that with no AF shell ($h_C \simeq 0.8$ K) due to the coupling between the AF and FM phases at the interface. The loop shifts obtained for the dots are small compared to those of spherical core/shell nanoparticles obtained in our previous studies [13].

In order to elucidate the origin of the low values of EB observed both experimentally and in the simulation, we have computed also the contribution of the shell interfacial (SI) spins to the hysteresis loop, since these are supposed to be responsible for the loop shift. The result, displayed in Fig. 4b, shows that a large fraction of the SI spins reverse with the magnetic field and, therefore, the local exchange field that acts on core spins to produce the EB comes only from the uncompensated magnetization of a small fraction of pinned spins. The small ratio of these spins to the core spins for nanodots of the sizes reported here is responsible for the reduced values of the $H_{EB}$ as compared to those reported for spherical core/shell nanoparticles, with typical sizes one order of magnitude smaller.

It is worth noticing also that, due to the cylindrical shape of the dots, the coordination of the SI spins is less than for a spherical particle, a fact that explains the biggest proportion of unpinned spins in dots than in spheres.

In summary, our results show that the thin CoO layer generated by oxygen exposition of lateral areas of Co nanodots is responsible for the low temperature enhancement of $H_C$ and the appearance of EB even for the biggest dots studied (110 nm). Monte Carlo simulations corroborate these results and give a proper explanation of the smaller values of EB compared with those observed in spherical core/shell nanoparticles, produced by the biggest ratio of the shell to core sizes in this case. The EB in the dots could be increased by an over-oxidation of the lateral area with ozone.

**ACKNOWLEDGEMENTS**

We acknowledge financial support from Spanish MEC (MAT2008-06765-C02-01/NAN, MAT2006-13572-C02-01, MAT2005-02601, MAT2006–03999, NAN2004-08805-CO4-01/02 and Consolider-Ingenio 2010 projects CSD2006–00012, CSD2007-00041 and CSD2008-00023) and European Commission (Phoremost NoE and NMP-FP7-214107 projects). As well as CSIC-PIF 200560F0120 project, FEDER program and Generalitat de Catalunya (2005SGR-00509), Comunidad de Madrid (S-0505/MAT/0194 NANOMAGNET ) and CESCA and CEPBA under coordination of C4 for computer facilities.

**FIGURES**

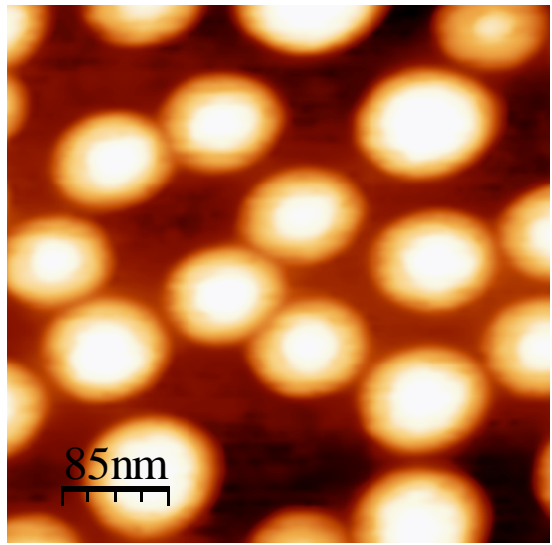

**Fig. 1** (Color online) AFM images of the Au/Co/Au nanopillars sample obtained with 60 nm diameter polystyrene sphere (N60).

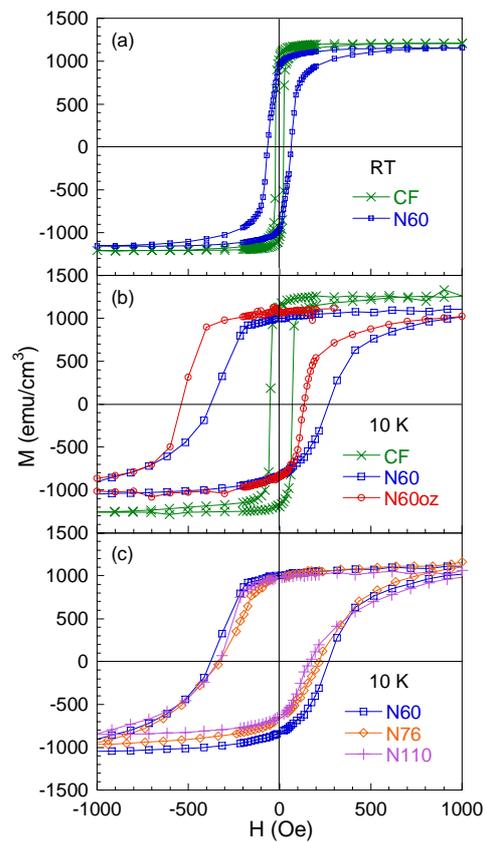

**Fig. 2** (Color online) (a) Magnetization curves measured at room temperature for the continuous (CF) and nanostructured film with 60 nm diameter dots (N60). (b) Magnetization curves at 10K after field cooling for the N60 sample before and after ozone attack. (c) Magnetization curves at 10K after field cooling for the three different nanostructured samples (N60, N76 and N110).

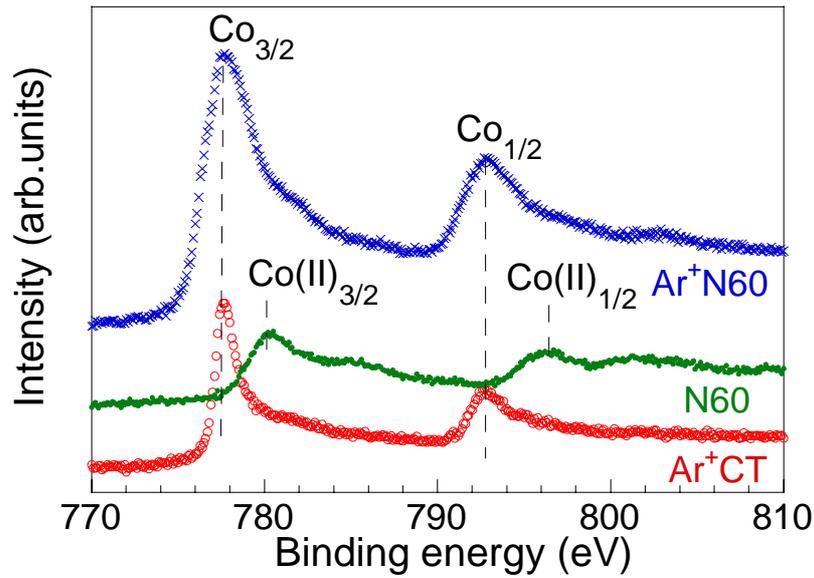

Fig. 3 (Color online) X-ray photoemission spectroscopy for the 60nm sample measured in the continuous film after small Ar$^+$ attack and in the nanostructured region before and after Ar$^+$ attack.

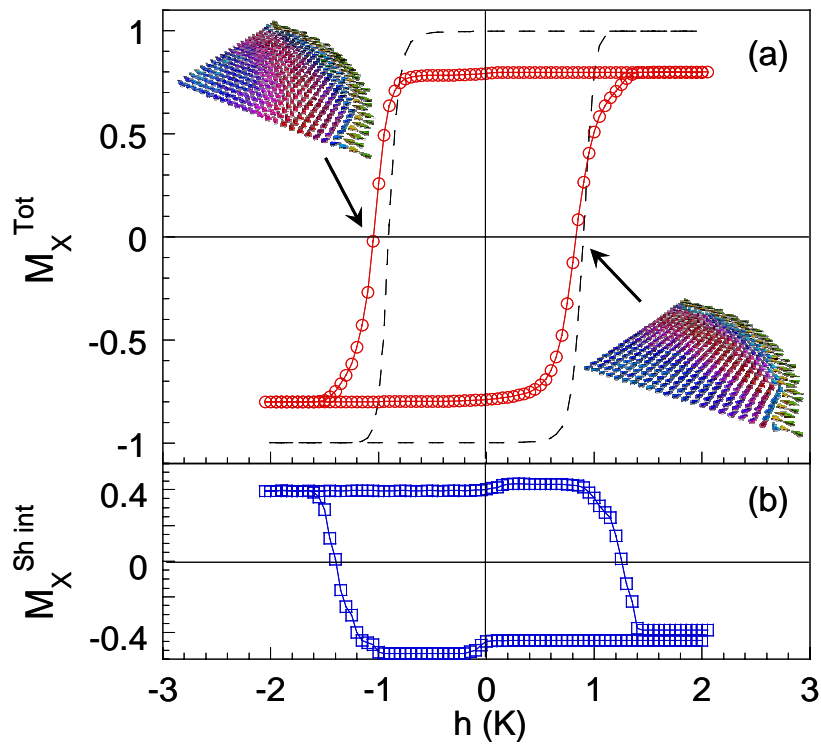

Fig. 4 (Color online) (a) Simulated hysteresis loop of a cylindrical dot of dimensions D'= 48 a and H'= 8 a and shell thickness $t_{Sh}$= 3 a obtained after field cooling from a high temperature T>TC down to 0.1 K in a magnetic field $h_{FC}$= 2 K (circles). Dashed lines correspond to a dot with the same dimensions but without the AF shell. Insets show the magnetic configuration of the core/shell dot at the coercive field points. Only a slice through a central plane of a quarter of the dot is displayed. Green (yellow) cones stand for the shell (shell interfacial) spins while the core spins have been colored from cyan to red according to the magnetization component transverse to the field. (b) Contribution of the AF shell spins at the interface of the core/shell dot with $t_{Sh}$= 3 a to the hysteresis loop shown above.